\documentclass[prl,showpacs,floatfix,twocolumn]{revtex4}

\usepackage{bm}
\usepackage{graphics}
\usepackage{graphicx}
\usepackage[dvips,usenames]{color}

\newcommand{\vs}{\vspace{0mm}}

\newcommand{\be}{\begin{equation}}
\newcommand{\ee}{\end{equation}}
\newcommand{\ba}{\begin{eqnarray}}
\newcommand{\ea}{\end{eqnarray}}
\newcommand{\NL}{\nonumber \\}

\newcommand{\VLASOV}[1]{\frac{\partial #1}{\partial t}}

\newcommand{\AuthorTeam}{
\author{T.~S.~Bir\'o\footnote{On leave from KFKI RMKI Budapest, Hungary}} 
\affiliation{
 Institute for Theoretical Physics, University of Giessen,
 D 35392 Heinrich-Buff-Ring 16, Giessen, Germany
}
\author{G.~Purcsel\footnote{On leave from KFKI RMKI Budapest, Hungary}} 
\affiliation{
 Institute for Theoretical Physics, University of Giessen,
 D 35392 Heinrich-Buff-Ring 16, Giessen, Germany
}
}

\begin{document}

\title{{ Non-extensive Boltzmann Equation and Hadronization}}

\AuthorTeam


\vs
\begin{abstract}
 For a general, associative addition rule defining a non-extensive thermodynamics
 we construct the strict monotonic function, which transforms it to a normal extensive
 quantity. We investigate the evolution of the one-particle distribution in the
 framework of a two-body Boltzmann equation supported with a non-extensive
 energy addition rule. An H-theorem can be proven for the extensive
 function of the non-extensive entropy. The equilibrium distribution
 is exponential in the extensive quantity associated to the original one-particle
 energy with the non-extensive addition rule.
 We propose that for describing the hadronization of quark matter non-extensive
 rules may apply.
\end{abstract}

\maketitle

\vs

\vs
There are several experimental evidences on power-law tailed statistical distributions
of single particle energy, momenta or velocity. In particular hadron transverse momentum
spectra at central rapidity, which stem from elementary particle and heavy ion 
collisions, can be well fitted by a formula reflecting 
$m_T$-scaling\cite{PHENIX,STAR,ZEUS,MAREK,BECK,TASSO,MT-SCALING,T-TEAM-FITS}:
\be
 f(p_T) \sim \left(1 + m_T/E_c \right)^{-v}.
 \label{TSALLIS-DISTR}
\ee
Interpreting these spectra as a distribution in the
transverse directions at zero rapidity, the single particle energy is given by
$E=m_T=\sqrt{p_T^2+m^2}$ for a relativistic particle with mass $m$
(in units setting the lightspeed to $c=1$).
Amazingly this formula describes exactly the Tsallis distribution $f(E)$, 
which was obtained by using theoretical arguments of thermodynamical 
nature\cite{TSALLIS-ENTROPY}. 
Distributions with power-law like tail, in particular the Tsallis distribution,
can be seen in many areas where statistical models 
apply\cite{MULT-NOISE,WILK,BIALAS,FLORKOW,KODAMA,RAF-BOLTZMANN-EQ,WAL-RAFELSKI,TSB-RHIC-SCHOOL,ASTRO}. 
It was investigated as a generic feature in the framework
of non-extensive thermodynamics\cite{TSALLIS-RULES,TSALLIS-WANG,TSALLIS-FOKKER}. 
Tsallis has suggested an expression for the entropy, 
also encountered earlier by others\cite{TSALLIS-ENTROPY,OTHER-ENTROPIES}, 
which would be an alternative to the Boltzmann formula. 
From this, with a canonical constraint on the total energy,
the distribution (\ref{TSALLIS-DISTR}) can be derived.
Without being able to exclude a non-equilibrium interpretation 
of the power-law tail, it is tempting
to investigate the possibility that some non-exponential spectra would be a result
of a particular form of equilibrium, featuring characteristics of a
non-extensive thermodynamics. 

\vs
In this paper we propose a possible way to understand power-law tailed
energy distributions as equilibrium solutions
to a slightly generalized two-body Boltzmann equation.
More specifically we show that this two-body Boltzmann-equation allows for 
non-exponential stationary single-particle distributions, if the two-body distribution 
{\em factorizes}, but the two-body energy is {\em not extensive}. 
The Tsallis distribution is a special case thereof. The pair-energy,
typical for generating this distribution, is a product of the
single particle energies.


\vs

\vs
It is a widespread belief that only the exponential distribution can be the stationary
solution to the Boltzmann equation, but this statement is true only with 
a few restrictions: i) if the two-particle
distributions factorize, ii) the two-particle energies are additive
in the single-particle energies ($E_{12}=E_1+E_2$ ) and 
iii) the collision rate is multilinear in the one-particle densities.

\vs
A generalization of the original Boltzmann equation has been pioneered by
Kaniadakis\cite{BOLTZMANN-GENERAL} considering a general, nonlinear
density dependence of the collision rates. An $''H_q''$ theorem for the particular
Tsallis form of the collision rate has been derived by Lima, Silva and 
Plastino\cite{HQ-THEOREM}.
Here we follow another ansatz, we modify the linear Boltzmann equation
in the energy balance part only: Instead of requiring
$E_1+E_2=E_3+E_4$ in a $1+2\leftrightarrow 3+4$ two-body collision we
consider a general, not necessarily extensive, rule:
\be
 h(E_1,E_2) \: = \: h(E_3,E_4).
\label{THE-RULE}
\ee
It is physically sensible to choose the function $h(x,y)$ symmetric and
satisfying $h(E,0)=h(0,E)=E$. Also, for applying the same rule for
subsystems combined themselves of subsystems, associativity is
required: $h(h(x,y),z)=h(x,h(y,z))$. This way the same rule applies for
the elementary two-particle system as for large subsystems in the
thermodynamical limit. It is known that the general mathematical solution
of the associativity requirement is given by
\be
 h(x,y) \: = \: X^{-1}\left( X(x) + X(y) \right),
\label{ASSOC-SOL}
\ee
with $X(0)=0$ and $X(t)$ being a continuous, strict monotonic 
function\cite{FUNC-EQ-MATH}.
Composing the formula (\ref{ASSOC-SOL})
with the function $X$ and taking the partial derivative with respect to
$y$ at $y=0$ one obtains an ordinary differential equation for $X(x)$
with the solution
\be
 X(E) \: = \: X'(0) \, \int_0^E \frac{dx}{\frac{\partial h}{\partial y}(x,0)}.
\label{QUASI-ENERGY}
\ee
Due to $X(h(E_1,E_2)) = X(E_1)+X(E_2)$, the quasi-energy $X(E)$ is
an additive quantity and the rule (\ref{THE-RULE}) is equivalent to
\be
 X(E_1) \, + \, X(E_2) \: = \: X(E_3) \, + \, X(E_4).
\label{ADD-RULE}
\ee
Applying such a general energy addition rule (\ref{ADD-RULE}),
the rate of change of the one-particle distribution is given by
\be
\VLASOV{f(p_1)}=\int_{p_2,p_3,p_4}\!\!\!\!\!\!\!\!\!\!\!\!\!\!\!{\cal W}(p_1,p_2;p_3,p_4)
\Delta \, \left[ f(p_3)f(p_4) - f(p_1)f(p_2) \right].
\label{BOLTZMANN-2-BODY}
\ee
with the symmetric transition probability ${\cal W}$ and the constraint
\be
\Delta \: = \: \delta^3(\vec{p}_1+\vec{p}_2-\vec{p}_3-\vec{p}_4) \,
	\delta\left(h(E_1,E_2) - h(E_3,E_4)\right).
\label{GEN-CONSTR}
\ee
In equilibrium the distributions depend on the phase space points
through the energy variables only and the detailed balance principle requires
\be
 f(E_1) \, f(E_2) \: = \: f(E_3) \, f(E_4).
\label{EQUIL}
\ee
With the generalized constraint (\ref{GEN-CONSTR}) 
this relation is satisfied by
\be
 f(E) \: = \: f(0) \exp(-X(E)/T)
\label{EQ-SOLUTION}
\ee
with $1/T=-f'(0)/f(0)$ and $X(E)$ given by eq.(\ref{QUASI-ENERGY}).
In the extensive case $h(x,y)=x+y$ leads to $X(E)=E$, for the 
Tsalis-type energy addition rule\cite{TSALLIS-RULES,TSALLIS-WANG}, 
\be
h(x,y) = x + y + a x y,
 \label{TS-ENERG}
\ee
one obtains $X(E) \: = \: \frac{1}{a} \ln(1+aE)$ and
\be
 f(E) \: = \: f(0) \left( 1 + aE \right)^{-1/aT}.
\ee
Since the energy addition rule (\ref{THE-RULE}) conserves the quantity
$h(E_1,E_2)$ in a microcollision, the new energies after the collision also lie
on the $h$=constant line. Due to the  additivity of the quasi-energy,
$X(E)$, the total sum $X_{tot}=\sum_i X(E_i)$, is a conserved quantity.
This rule can be applied in numerical simulations, too.
Since our present goal is to find the equilibrium
only, we may assume constant transition probabilities.

\begin{figure}
\begin{center}
 \includegraphics[width=0.20\textwidth,angle=-90]{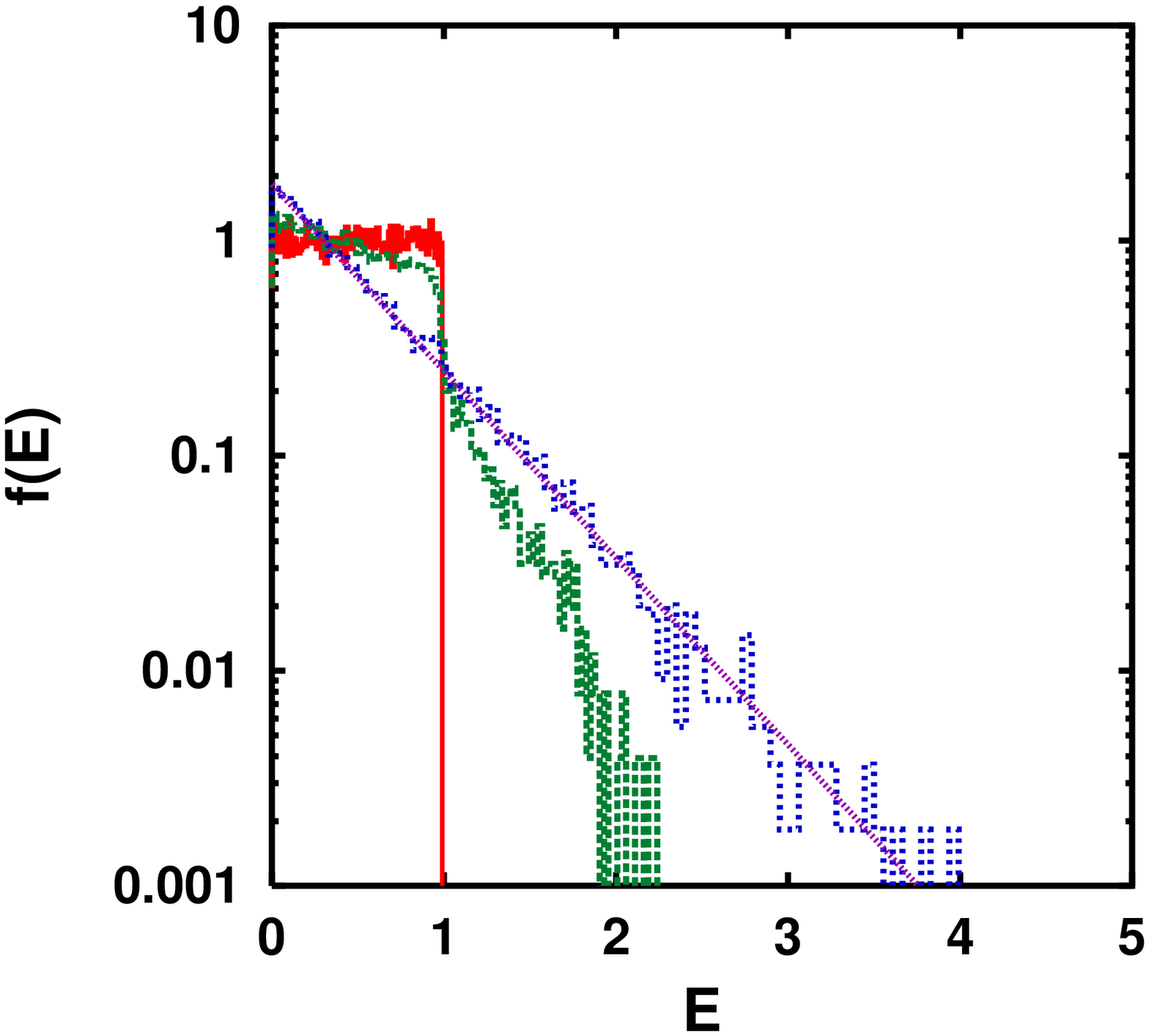}\includegraphics[width=0.20\textwidth,angle=-90]{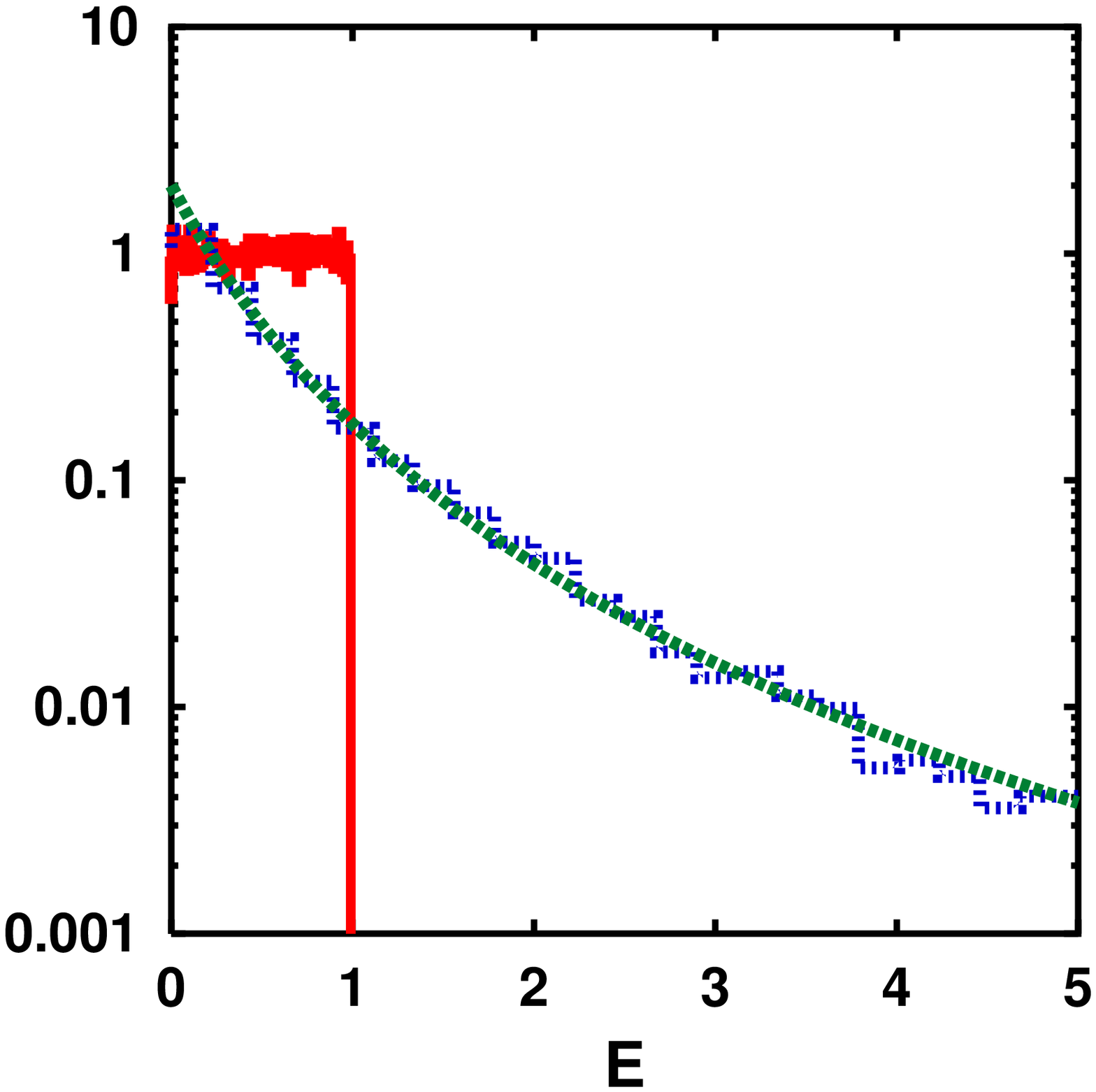}
 \end{center}
 \caption{
 \label{FIG-TS}
 	(Color online)
 	Evolution towards the Boltzmann distribution 
	for $a=0$ (left part) and towards the
	Tsallis one for $a=1$ (right part) using $h(E_1,E_2)=E_1+E_2+aE_1E_2$.
	For detailed explanation see text.
 }
 \end{figure}

\vs
Fig.\ref{FIG-TS} presents results of a simple test 
particle simulation with the rule (\ref{TS-ENERG}).
We mostly started with a uniform energy distribution between
zero and $E_0=1$ with a fixed number of particles $N=10^4$ 
(red, full line)\footnote{We also have tested several other initial distributions
not discussed here.}.
The one-particle energy distribution evolves towards the
well-known exponential curve for $a=0$,
shown in the left part of Fig.\ref{FIG-TS}. This snapshot
was taken after $200$ two-body collisions per particle
(blue, short dashed line). The analytical fit to this histogram is 
given by $2 e^{-E/T}$ (with $T = E_0/2 = 0.5$ in this case).
An intermediate stage of the evolution after $0.4$ collisions per 
particle is also plotted in this figure (green, long dashed line).
Using the prescription (\ref{TS-ENERG}) with $a=1$, the stationary
solution becomes a Tsallis distribution. 
The numerical evolution from the uniform energy distribution can be inspected
in the right side of Fig.\ref{FIG-TS}. 
The fit to the final curve is given by 
$f(E)=2.6 (1+aE)^{-3.6}$. All distributions are normalized to one.

\vs

\vs
It is in order to make some remark on the energy conservation. 
For $h(x,y)=x+y$ we simulate a closed system with elastic collisions:
The sum, $U=\sum_{i=1}^N E_i$, does not change in any of the binary
collisions.  The situation changes by using a non-extensive formula 
for $h(x,y)$, like in the case of the Tsallis prescription. 
With a constant positive (negative) $a$,
the bare energy sum, $U$, is decreasing (increasing) 
while approaching the stationary distribution.
This is typical for open systems gaining or loosing energy during their
evolution towards a stationary state.

\vs 
One may incline to consider the conserved quasi-energy, $X(E)$, 
as an in-medium one particle energy. 
The interesting point is that in general any prescription, $h(E_1,E_2)$
-- defining a version of the non-extensive thermodynamics --, 
is equivalent to considering a quasi-energy, $X(E)$. 
For small energies one expects a restoration of the
extensive rule and $X(E)\approx E$, \hbox{$X'(0)=1$.}
Whenever the pair energy is repulsive (attractive), 
$h(E_1,E_2) \ge E_1+E_2$ ($h(E_1,E_2) \le E_1+E_2$), 
a rising quasi-energy is smaller (bigger) than the free one, 
$X(E) \le E$ ($X(E) \ge E$)\footnote{For rising quasi-energy $X(x)$ 
and for $h(x,y)\ge x+y$ we have $X(x)+X(y)=X(h(x,y)) \ge X(x+y)$. 
Expanding for small $y$ at arbitrary $x$
it gives $X'(0)\ge X'(x)$. With $X(0)=0$ and $X'(0)=1$ 
the result $X(x)\le x$ follows.}.
This leads to a tail of the stationary distribution in the free single
particle energy, $f(E)$, which is above (below) the exponential curve. 
This phenomenon is hard to distinguish from a power-law tail 
numerically\footnote{The correct analysis of experimental data should inspect the sliding inverse logarithmic slope as a function of the single particle energy. Its deviation from linear is also
a deviation from the Tsallis statistics and from the power law.}.

\vs

\vs
The question arises that -- constrained by the conserved number of particles,
$N = \int f \, d\Gamma$, 
and the total quasi-energy, $X_{{\rm tot}} \, = \, \int f\, X(E) \, d\Gamma$,
 -- what is the proper formula for the entropy which  grows when
approaching the stationary distribution. If the addition rule of the
non-extensive entropy, $s$, is given by $h_s(x,y)$, then the quasi-entropy,
$X_s(s)$, is additive and the total entropy is given by
\be
X_s(S_{tot})\: = \: \int\! f \, X_s(s(f)) \, d\Gamma = \int\!\sigma(f)\, d\Gamma. 
\label{TOT-QUASI-ENTROPY}
\ee
Its rate of change,
$\dot{X}_s(S_{tot}) \: = \: \int \dot{f} \, \sigma'(f) \, d\Gamma$ can be expressed with
the help of the Boltzmann equation (\ref{BOLTZMANN-2-BODY}).

\begin{figure}
\begin{center}
 \includegraphics[width=0.22\textwidth,angle=-90]{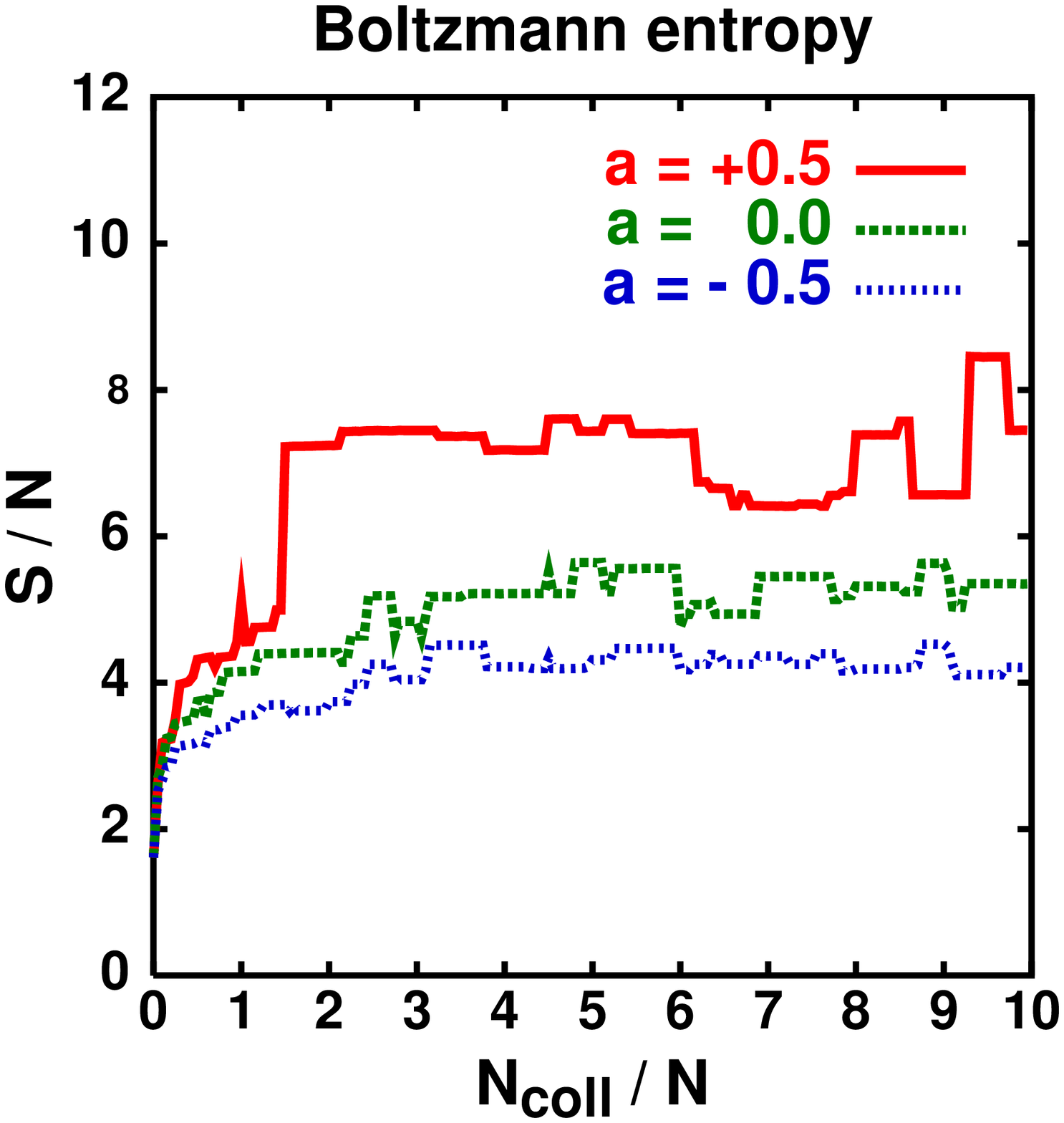}\includegraphics[width=0.22\textwidth,angle=-90]{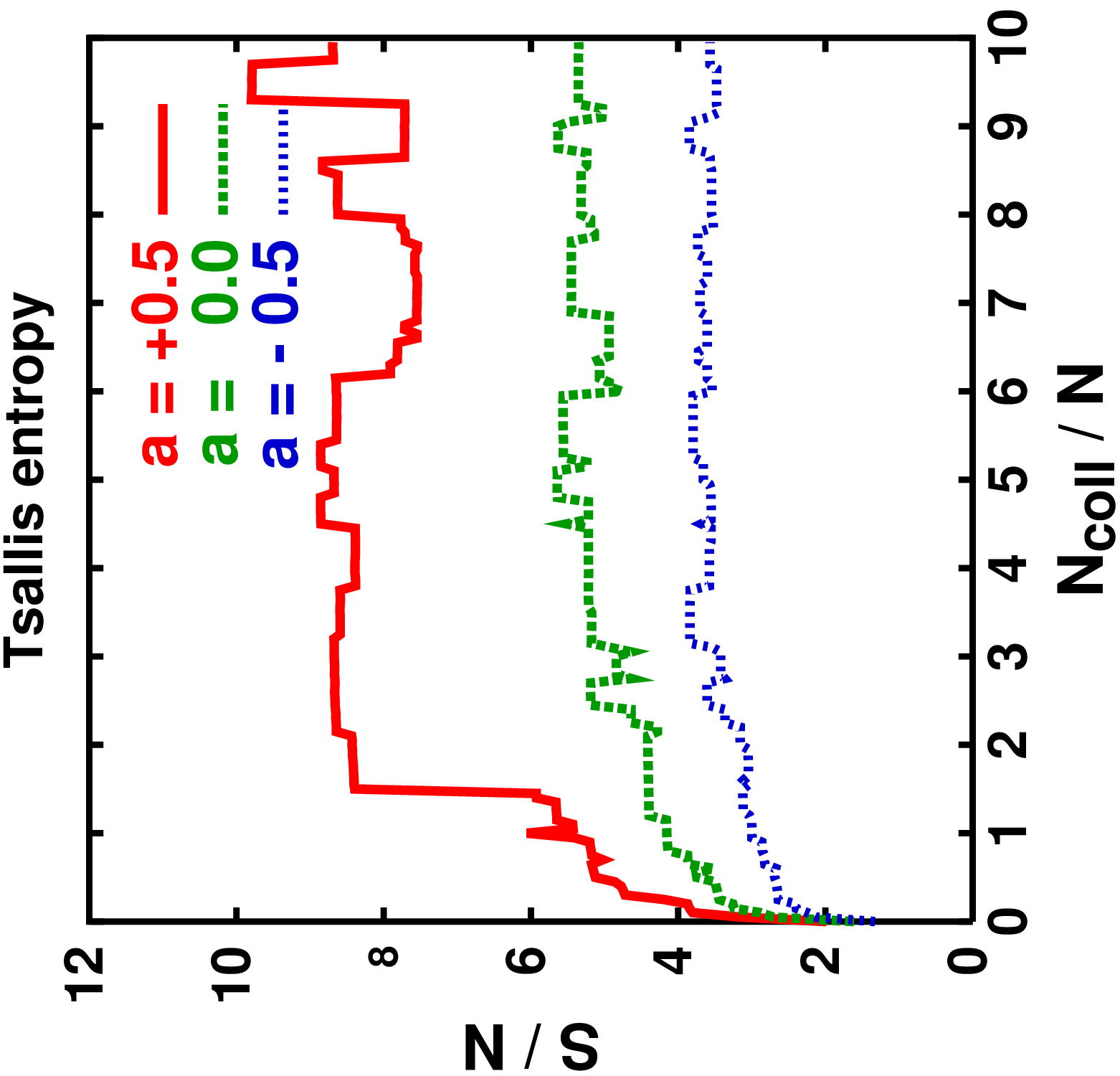}
 \end{center}
 \caption{ \label{FIG-ENTROPY} 
 (Color online) 
 Evolution of the Boltzmann and Tsallis entropies
 by applying the energy addition rule  $h(E_1,E_2)=E_1+E_2+aE_1E_2$ with different
 values of the parameter $a$.
 }
 \end{figure}

\vs
Assuming the symmetry properties $1\leftrightarrow 2$, $3\leftrightarrow 4$
and $(12)\leftrightarrow (34)$ for the constrained rate factor $w_{1234}={\cal W}\Delta$,
one easily derives
\ba
X_s'(S_{tot}) \: \dot{S}_{tot} \: = \: \frac{1}{4} \int_{1234} \!\!
	w_{1234} \left( f_3f_4 - f_1f_2 \right) \, \times & \, & \NL
	\left(\sigma'(f_1)+\sigma'(f_2)-\sigma'(f_3)-\sigma'(f_4) \right). & \, &
\label{H-theorem}	
\ea
A definite sign for this quantity can be obtained  due to the additivity of
$X_s(s(f))=\sigma(f)/f$, leading to the unique solution $X_s(s(f))=B\ln f$
\footnote{From $\sigma(f_1)/f_1+\sigma(f_2)/f_2=\sigma(f_1f_2)/(f_1f_2)$
it is easy to derive that 
$\sigma'(f_1)+\sigma'(f_2)=2\sigma'(f_1f_2)-\sigma(f_1f_2)/(f_1f_2)$
is a function of $f_1f_2$ only.
The general solution of the functional equation
$\sigma'(f_1)+\sigma'(f_2)=\Phi(f_1f_2)$ 
is given by  $\Phi(z)=2A+B\ln z$, and $\sigma'(x)=A+B\ln x$ with $z=xy$.
Satisfying $\Phi(z)=2\sigma'(z)-\sigma(z)/z$ is possible by $A=B$ only.
}
.
For $B=-k_B$ (Boltzmann's constant) $\dot{S}_{tot} \ge 0$ follows.
Using the $k_B=1$ unit system we arrive at $X_s(s(f))=-\ln f$,
and the expression for the total additive quasi-entropy (\ref{TOT-QUASI-ENTROPY}),
coincides with Boltzmann's original suggestion. At the same time,
applying a non-extensive addition rule for the entropy,
$h_s(x,y)=x+y+(1-q)xy$, as Tsallis did, we have $X_s(s)=\frac{1}{1-q}\ln(1+(1-q)s)$
(Abe's formula\cite{ABE}), and from $X_s(s)=-\ln f $
one obtains Tsallis' entropy formula
\be
  f s(f) = \frac{f^q-f}{1-q}.
\ee
In numerical simulations 
we observe that the Tsallis-type non-extensive energy addition rule, applied to
a test-particle simulation with two-body collisions, maximizes
the Boltzmann-quasi-entropy (cf. Fig.\ref{FIG-ENTROPY}).
The stationary distribution is nevertheless a Tsallis distribution (cf. Fig.\ref{FIG-TS}).

\vs
It is interesting to note that, as in many papers\cite{TSALLIS-RULES}, 
applying a canonical constraint on the bare total energy,
$U = \int\!E f(E) \, d\Gamma$, when seeking for the equilibrium distribution 
one obtains equivalent results for the Tsallis case. 
Instead of using the (due to the H-theorem guaranteed correct)
formula, 
\be
\frac{d}{df}\left(fX_s(s(f))\right) \: = \: \alpha + \beta X(E),
\label{CORRECT-CANONICAL}
\ee
where $X_s(s)$ and $X(E)$ are both Tsallis-like expressions,
the naive assumption of
\be
\frac{d}{df}\left(fs(f)\right) \: = \: \alpha + \beta E,
\label{NAIV-CANONICAL}
\ee
also leads also to a power-law distribution.
In the first case $X_s(s(f))=-\ln f$ independently of the non-extensive rule
definition for the entropy, and the equilibrium solution is 
$f_{eq}\propto \exp(-X(E)/T) = (1+aE)^{-1/(aT)}$ with $\beta=1/T$.
In the second case the particular form of $s(f)$ is used and the non-extensive
addition rule for the energy is ignored. (This case has nothing to do with our 
simulation.) The distribution resulting from (\ref{NAIV-CANONICAL})
is $f_{eq}\propto (1+aE/(1+\alpha aE))^{-1/(aT)}$ when $q=1-aT$.

\vs
In numerical simulations both the Boltzmann and the Tsallis entropy 
increase,  disregarding the fluctuations 
due to the finite number of test particles, $N=10^4$, and due to finite binning
resolution in energy, ($N_{bin}=100$ while the maximal energy, $E_{max}$ is changing).
The increase is a trend, it is not rigorously fulfilled in each microscopic
collision; it is a common behavior of molecular dynamical simulations.
The ratio of the two entropy expressions
also fluctuates somewhat, but the Tsallis entropy in the repulsive (attractive)
case clearly stays bigger (smaller) than the Boltzmann entropy.

\vs

\vs
Physical realizations of non-extensive systems may be discovered depending on
our knowledge about the microscopical forces
influencing the particles during the pair-interactions. For such forces being
repulsive, the canonical one-particle energy distribution has a tail above the
exponential curve, for attractive interactions below. As a rule, as long as this
modification is small, it is extremely difficult to see the non-exponential
tail in the bare one-particle energy distribution both in experiments and
in numerical simulations. Such power-law tails are prominent in elementary
particle spectra in high energy experiments, but their traditional explanation
does not assume an equilibrium state.

\vs
In the quark-gluon plasma (QGP), or more generally in a parton matter before
hadronization color non-singlet objects are the single particles.
Eventually all form hadrons, in the soft sector perhaps by recombination
and in the hard sector dominantly by fragmentation. In both cases 
a long-range interaction between color non-singlet partons, connected to the
physical phenomenon confinement, is present in the background. 
In the following we consider a simple model for including this type of
non-perturbative pair-interaction.

\vs
For the sake of simplicity let us restrict ourselves to two body processes
between color triplets and anti-triplets. This is the most common way of
meson formation. It is also an important part of baryon formation 
due to quark - diquark fusion. The pairs of such partons,
while they constantly interact, are either in a color singlet or in a color octet state
(in a QGP in one case from nine a singlet, otherwise an octet). The energy
of the two-parton system is given by
\be
 E^{{\rm \: color \, state}}_{12} = E_1 + E_2 + \Delta^{{\rm color \, state}},
\label{TWO-PARTONS}
\ee
where the singlet channel should be attractive (relative to the free partons).
The color average is supposed to be vanishing,
$ \Delta^{{\rm singlet}} + 8 \Delta^{{\rm octet}} = 0$.
This is certainly the case for interactions like in the Heisenberg-model
of magnets, where the pair-potential
is proportional to the product of symmetry generators in the corresponding
spin representation.  For SU(3) color this is also the case.
The singlet charge is zero, the octet charge square is $Q^2_{{\rm octet}} = 3$.
The triplet and anti-triplet both have $Q^2_{{\rm triplet}} = 4/3$. The 
Heisenberg-magnet-like
interaction in color has therefore a factor of $-8/3$ for the singlet and a factor
of $3-8/3=1/3$ for the octet. Their degeneracy-weighted sum is zero.

\vs
For considering the possibility of a non-Boltzmann distribution in quark
matter we further assume a Coulomb-like interaction. In this case
$ \Delta^{{\rm singlet}} \: = \:  2E_{12}^{{\rm \: rel.kin.}}$
from the binding in the color singlet channel. 
For the search after a stationary single-quark distribution of the two-body
Boltzmann equation in the octet channel it accounts to consider,
\be
 E_{12}^{{\rm \: octet}} \, = \, 
	E_1 \: + \: E_2 \: + \: \frac{1}{4} \: E_{12}^{{\rm \: rel.kin.}}.
 \label{OCTET}
\ee
The rest is kinematical consideration. We assume the coalescence of two massless
partons to a (nearly) massless hadron.
Due to the triangle inequality, the kinetic energy of the relative motion of
two massless partons is non-negative,
\be
 E_{12}^{{\rm \: rel.kin.}} = |\vec{p}_1+\vec{p_2}| - |\vec{p}_1| - |\vec{p}_2| \ge 0.
\ee
For small relative angles between the momentum vectors, $\vartheta$, this is approximated by
\be
 E_{12}^{{\rm \: rel.kin.}} = \frac{2 E_1E_2\,\sin^2(\vartheta/2)}{E_1+E_2} 
 \label{KINETIC}
\ee
The sum of the individual parton energies in the same approximation is close to
$E_1+E_2\approx P = |\vec{p}_1+\vec{p}_2|$.
Even very hard partons with a high value of the total pair momentum,
$P$, need a little relative motion for interacting:
in the singlet channel to eventually form hadrons, in the octet channel to
maintain a single-particle quark-distribution typical for the pre-hadronic phase.
The stationary version of this distribution, while detailed balance is
satisfied on the two-body level, is often found to be close to the Tsallis distribution.
We propose that the above mechanism, from the comparison of 
eqs.(\ref{TS-ENERG}), (\ref{OCTET}) and (\ref{KINETIC}) leading to
\be
 E_c = 1/a = \frac{2P}{\sin^2(\vartheta/2)},
\ee
may be in the background of such findings. Asymptotical freedom is recovered
as for very fast partons $E_c\rightarrow \infty$ with $P \rightarrow\infty$,
and so the one particle energies of a colliding pair become additive.


\vs
In conclusion we have investigated deterministic, non-extensive energy
addition rules in two-body collisions. We have pointed out that instead of
the one-particle energy a quasi-energy is conserved by such rules in each
collision, leading to a non-Boltzmannian stationary distribution in the bare one-particle
energy. In particular the Tsallis distribution is obtained by using a 
Tsallis-type non-extensive energy addition rule. The corresponding 
conserved quasi-energy is identical to that proposed by Q.Wang\cite{TSALLIS-WANG}.
Modifications to the extensive energy addition rule may have to be considered if there
is a statistically important pair-interaction between particles. The stationary
canonical distribution becomes exponential in the conserved quasi-energy $X(E)$,
but it is non-exponential in terms of the free particle energy $E$.
The Boltzmann entropy,  $S_B=X_s(S_{tot})=-\int f \ln f d\Gamma$,
is never decreasing and reaches its maximum  at this distribution.
Alternative expressions for the entropy, in particular the one 
promoted by Tsallis, correspond to a non-extensive entropy addition rule
which defines $X_s(s)$.
Notably, in the Tsallis case also the naive integral, $S_T=\int f s(f) d\Gamma$,
seem to increase in numerical simulations as the elementary collisions
proceed. The Tsallis-type
energy and entropy addition rules are correlated by $q=1-aT$.

\vs
As a possible physical realization 
we have proposed a mechanism leading to nearly Tsallis-distributed quarks
in quark matter and hadrons which eventually form. This mechanism considers
a color state dependent pair-energy 
based on general arguments describing a quantum symmetry. The essential clue
leading to our result then hides in the use of a virial theorem which
connects the color interaction with kinematical factors of the quark
pair. In a certain approximation the modification of the familiar two-body 
energy conservation factor in the Boltzmann equation receives 
a term proportional to the product of single-quark kinetic energies to leading
order in the ultrarelativistic expansion. 
The Tsallis distribution turns out to be an approximation
next simplest to the original Boltzmannian 
one\footnote{Due to (\ref{ASSOC-SOL}) any small-energy expansion 
$X(E)=E-aE^2/2+\ldots$ leads to the rule $h(x,y)=x+y+axy$.}.

\vs
Our picture naturally connects the processes maintaining a possible stationary
distribution among colored partons with the hadronization process. In the
approximation discussed in this paper the power of the power-law tail in
hadronic spectra equals to the power occurring in the single-quark distribution
in quark matter. As a consequence mesonic and baryonic powers are also equal
to each other. This agrees with experimental findings well, although  the 
recombination assumption predicting a baryonic to mesonic power ratio of $3:2$
also cannot be excluded, when considering the relatively high error bars. 
This result is, however, simpler
and more generic. Here only a balance between kinetic and potential energy
in the relative motion of quarks has been assumed besides some basic color
properties of the pairwise interaction.


\vs
{\bf Acknowledgment}

Enlightening discussions with Prof. Carsten Greiner and Dr.
Zhe Xu at the University of Frankfurt as well as with
Drs. G\'eza Gy\"orgyi at E\"otv\"os University
and Antal Jakov\'ac at the Technical University Budapest are hereby
gratefully acknowledged.
This work has been supported by the Hungarian National Science Fund OTKA
(T034269, T49466) 
and the Deutsche Forschungsgemeinschaft due to a Mercator
Professorship for T.S.B.


%
%


\end{document}